\documentclass[12pt]{article}

\usepackage{amsmath}
\usepackage{amssymb}
\usepackage{cite}

\def\da{{\dot{a}}}
\def\db{{\dot{b}}}

\def\nn{\nonumber}

\numberwithin{equation}{section}

\title{Cubic interactions of $d4$ irreducible massless\\
 higher spin fields within BRST approach}

\author{I.L. Buchbinder${}^{abc}$\thanks{joseph@tspu.edu.ru}, V.A. Krykhtin${}^{ab}$\thanks{krykhtin@tspu.edu.ru}
T.V. Snegirev${}^{ab}$\thanks{snegirev@tspu.edu.ru}
\\[0.5cm]
\it{\small ${}^a$Center of Theoretical Physics, Tomsk State
Pedagogical University,}\\
\it{\small Tomsk, 634061, Russia}\\
\it{\small ${}^b$National Research Tomsk State University, Tomsk
634050, Russia}\\
\it{\small ${}^c$Lab. for Theor. Cosmology, International Center
of}\\
\it{\small Gravity and Cosmos, Tomsk State University
of}\\
\it{\small Control Systems and Radioelectronics, 634050 Tomsk,
Russia }}

\date{}

\begin{document}

\maketitle

\begin{abstract}
We develop an approach to constructing the manifestly Lorentz
covariant cubic interaction vertices for the four-dimensional
massless higher spin bosonic fields with two-component dotted and
undotted spinor indices. Such fields automatically satisfy the
traceless conditions what simplify form of the equations determining
the irreducible massless representation of the Poincar\'e group with
given helicity. The cubic vertex is formulated in the framework of
the BRST approach to higher spin field theory. Use of the above
spin-tensor fields allows to simplify a form of the BRST-charge and
hence to find the cubic vertices just in terms of irreducible higher
spin fields. We derive an equation for the cubic vertex and find
solutions for arbitrary spins $s_1,s_2,s_3$ with the number of
derivatives $s_1+s_2+s_3$ in the vertex. As an example, we
explicitly construct a vertex corresponding to the interaction of a
higher spin field with scalars.

\end{abstract}

\thispagestyle{empty}
\newpage
\setcounter{page}{1}

\section{Introduction}
Study the various aspects of higher spin filed theory still attracts
an attention motivated among other things by certain possibilities
for the development of new principles for constructing unified
models of fundamental interactions, including quantum gravity (see
e.g. the reviews \cite{BBS,BHV,BBCCFGSS22,DP} and the reference
therein for current progress).

Cubic interactions are the first approximation in the theory of
interacting fields. Their peculiarity is that the cubic interaction
for given three fields does not depend on the presence or
absence of any other fields in the full nonlinear theory. Thus, they
are model independent and can be classified. The complete
classification of consistent cubic interaction vertices of massless
and massive fields of arbitrary spin was constructed in the
light-cone formalism in the space dimensions $d\geq4$ by Metsaev
\cite{Met05,Met07b,Met18a}. In the simplest case of three massless
symmetric fields with spins $s_1$, $s_2$, $s_3$ in Minkowsky space,
the cubic vertices are characterized by the number of derivatives
$k$
\begin{eqnarray*}
k_{min}=s_1+s_2+s_3-2s_{min}\leq k\leq s_1+s_2+s_3=k_{max},\quad d>4.
\end{eqnarray*}
There are only two vertices in four dimensions $d=4$  with
$k=k_{min},k_{max}$ (see e.g. \cite{DP}). Already in this case the
Lorentz-covariant realization of cubic interaction vertices for
higher spins requires very cumbersome calculations.

As is known the  irreducible massless higher spin representation of
the Poincar\'e group with integer spin $s$ is described by
metric-like symmetric tensor fields $\phi^{\mu_1...\mu_s}$ which
satisfy equation $\partial^2\phi^{\mu_1...\mu_s}=0$ and subject
transversality $\partial_\nu\phi^{\mu_1...\mu_{s-1}\nu}=0$ and
tracelessness $\phi^{\mu_1...\mu_{s-2}\nu}{}_\nu=0$ constraints. One
can expect that in the case of interacting fields these equation and
constraints should be modified in some way. One of the generic
problems in higher spin field theory is to derive these modified
equations and the corresponding constraints within the Lagrangian
formulation. This problem has been studied by many authors using
different approaches (see e.g.
\cite{BBL87,BFPT06,BL06,Zin08,BLS08,BBL10,FT10,MMR10a,MMR10b,JT11,Metsaev12,JLT12a,JLT12b,JT13,HGR13,CJM16,FMM16,KMP19,
FKM19,JT19,BKTM21} and the references therein). Also it is worth
noting the papers, where the consistent cubic vertices for massless
higher spin fields was constructed in the the frame-like formalism
\cite{Vas11,BPS12,ZinKh20}\footnote{Recently, there was developed a
general approach to gauge invariant deformations of gauge systems
\cite{BL1,BL2,BL3,BL4}, that opens the new possibilities to
construct the higher spin field vertices \cite{L1,L2}. We also point
out a new approach to problem of locality in the higher spin field
theory that can be related with a structure of the interaction
vertices \cite{Vas22}.}.

Recently the cubic vertices for symmetric massless higher spin
fields with integer spins were considered within the BRST approach
for $d\geq4$ \cite{BR21} in Minkowski space where the BRST charge
takes into account all the constraints. As it turned out, taking
into account the  of traceless constraints, although it requires
overcoming some formal difficulties, can lead to the appearance of
new terms in the cubic vertex in comparison with the formulation
without using the conditions of tracelessness in the BRST charge. In
this paper, we consider the same problem but from a different angle.

We want to pay attention that in $d4$ all the worries with
tracelessness constraints can be avoid if to work within of the
two-component spinor formalism. Irreducible massless fields with
integer spin $s$ are described in this case by multispinors
$\phi^{a_1...a_s\da_1...\da_s}$ satisfying the equation
$\partial^2\phi^{a_1...a_s\da_1...\da_s}=0$ and subject to
transversality condition
$\partial_{b\db}\phi^{a_1...a_{s-1}b\da_1...\da_{s-1}\db}=0$. There
are no need to use the tracelessness constraint. Our aim is to apply
the BRST approach for construction of cubic interaction for massless
bosonic fields in terms of spin-tensor fields with two component
dotted and undotted spinor indices.

The paper is organized as follows. In Section~\ref{sec:FreeBRST} we
present the basic aspects of the BRST approach for constructing the
free Lagrangian formulations of irreducible massless fields with an
arbitrary integer spin in two-component formalism. In Section 3 we
describes the general procedure of the BRST approach for
constructing interactions and present the BRST-closed condition for
cubic vertices. Particular solution to this condition with the
number of derivatives $k_{max}=s_1+s_2+s_3$ in the vertex.is given
in Section 4. One example of the interaction of a higher spin field
with scalars in more detail is discussed in Section 5. In section 6
we summarize the results obtained.

\section{BRST charge and free Lagrangian}\label{sec:FreeBRST}
In space of $4d$ multispinor tensors the  irreducible massless
higher spin fields with integer spin $s$ can be described by
fields $\phi_{a(s)\da(s)}$ subjected to constraints
\begin{equation}\label{IrConstraints}
\partial^{b\db}\phi_{a(s-1)b\da(s-1)\db}=0,\qquad
\partial^2\phi_{a(s)\da(s)}=0,
\end{equation}
where
$\partial^2=\partial_\mu\partial^\mu=-\frac12\partial_{a\da}\partial^{a\da}$.

In the framework of the BRST approach, the higher spin fields appear
as the coefficients in the vectors of the Fock space
\begin{eqnarray}\label{FockV}
|\phi\rangle=\sum_{s=0}^{\infty}|\phi^{(s)}\rangle,
&\quad&
|\phi^{(s)}\rangle=\frac{1}{s!}\phi_{a(s)}{}^{\da(s)}c{}^{a(s)}c{}_{\da(s)}|0\rangle
\,,
\\
&&
c{}^{a(s)}:= c^{a_1}\ldots c^{a_s}
\quad
c{}_{\da(s)}:=c{}_{\da_1}\ldots c{}_{\da_s}
\end{eqnarray}
generated by creation $c^a,c^\da$ and annihilation operators
$a^a,a^\da$
\begin{equation}
\langle0|c{}^a=\langle0|c{}^\da=0,\qquad
a^a|0\rangle={a}^\da|0\rangle=0,\qquad \langle0|0\rangle=1
\end{equation}
with following nonzero commutation relations
\begin{equation}
[a^a,c{}^b]=\varepsilon^{ab},\qquad
[a^\da,c{}^\db]=-\varepsilon^{\da\db}.
\end{equation}
The Hermitian conjugation in a Fock space is defined as follows
\begin{equation}
(a^a)^+={c}^{\da},\quad
({a}^{\da})^+=c{}^{a},\quad(c^a)^+={a}^{\da},\quad
({c}^{\da})^+=a{}^{a}.
\end{equation}

Let us introduce operators
\begin{equation}\label{Operators}
p^2=\partial^2,\quad l=a^a{a}^\da p_{a\da},\quad l^+=-c{}^ac^\da
p_{a\da},\quad p_{a\da}=\partial_{a\da},
\end{equation}
which act on the state of the Fock space as
\begin{eqnarray*}
p^2|\phi^{(s)}\rangle
&=&
\frac{1}{s!}\partial^2\phi_{a(s)}{}^{\da(s)}c{}^{a(s)}c{}_{\da(s)}|0\rangle
\,,
\\
l|\phi^{(s)}\rangle&=&-\frac{s^2}{s!}
\partial^a{}_{\da}\phi_{a(s)}{}^{\da(s)} c^{a(s-1)}c{}_{\da(s-1)}|0\rangle
\,,
\\
l^+|\phi^{(s)}\rangle&=&\frac{1}{s!}\partial_{b}{}^{\db}\phi_{a(s)}{}^{\da(s)}c{}^{a(s)b}c{}_{\da(s)\db}|0\rangle
\,.
\end{eqnarray*}
The constraints (\ref{IrConstraints}) in terms of operators
(\ref{Operators}) take form
\begin{equation}\label{PhysSt}
p^2|\phi^{(s)}\rangle=0,\qquad l|\phi^{(s)}\rangle=0.
\end{equation}

Note the set of operators $F_A=\{p^2,l,l^+\}$ is invariant under
Hermitian conjugation with respect to the scalar product
\begin{equation}
\langle\phi||\phi\rangle
\end{equation}
and form a closed algebra $[F_A,F_B]=f_{AB}{}^CF_C$ with the only
nonzero commutation relation
\begin{equation}
[l^+,l]=(N +\bar{N}+2)p^2,
\end{equation}
where
$$
N=c{}^a a_a,
\qquad
\bar{N}=c{}_{\da}{a}^{\da},
\qquad
(N)^+=\bar{N}.
$$

The Hermitian nilpotent BRST charge is constructed in the form
\begin{equation}\label{BRST_Charge_F}
Q=\eta^AF_A-\frac12\eta^A\eta^Bf_{AB}{}^C{\cal P}_C,
\quad
Q^+=Q,
\quad
Q^2=0,
\end{equation}
where $\eta_A=\{\theta, c^+,c\}$  and ${\cal P}_A=\{\pi, b, b^+\}$
are the fermionic ghosts and corresponding momenta (antighosts)
satisfying the anticommutation relations $\{\eta_A,{\cal
P}_B\}=\delta_{AB}$. The ghost and antighost variables have ghost
numbers $gh(\eta_A)=-gh({\cal P}_A)=1$ while $gh(F_A)=0$. Applying
formula (\ref{BRST_Charge_F}) we obtain the explicit expression for
BRST charge
\begin{equation}
\label{Q-final}
 Q=\theta p^2+c^+l+cl^++c^+c(N +\bar{N}+2)\pi
\,.
\end{equation}

The operator $Q$ (\ref{Q-final}) is nilpotent $Q^2=0$ and acts in the extended Fock space of the vectors
$|\Phi\rangle$ including dependence on the fermionic ghosts defined above.

We note here that there are two operators
\begin{equation}
\label{SbarS}
S=c{}^a a_a+c^+b+b^+c,
\qquad
\bar{S}=c{}_{\da}{a}^{\da}+c^+b+b^+c,
\end{equation}
that commute with the BRST operator \eqref{Q-final}.
Thus if we want to get Lagrangian for bosonic spin $s$ field \eqref{FockV} the state vector $|\Phi\rangle$  in the extended Fock space must satisfy
\begin{equation}
S|\Phi\rangle=s|\Phi\rangle,
\qquad
\bar{S}|\Phi\rangle=s|\Phi\rangle.
\end{equation}

The basic equation of motion in the BRST approach is postulated in the extended Fock space as follows
\begin{equation}
\label{basic-equation}
Q|\Phi\rangle=0.
\end{equation}
To reproduce the conditions of the irreducible representation (\ref{PhysSt})
from the equation of motion (\ref{basic-equation}) we define the vacuum of extended Fock space as
\begin{equation}
c|0\rangle=b|0\rangle=\pi|0\rangle=0
\end{equation}
and require that the ghost number of vector $|\Phi\rangle$ is equal
to the ghost number of initial vector (\ref{FockV}),
$gh(|\Phi\rangle)=gh(|\phi^{(s)}\rangle)=0$. Then the most general vector
$|\Phi\rangle$ of the extended Fock space has the form
\begin{equation}\label{FockTV}
|\Phi\rangle=|\phi^{(s)}\rangle+\theta b^+|\phi_{1}^{(s-1)}\rangle+c^+
b^+|\phi_{2}^{(s-2)}\rangle
\end{equation}
where the $|\phi^{(s)}\rangle$ is the initial vector (\ref{FockV}) and
the $|\phi_1^{(s-1)}\rangle, |\phi_2^{(s-2)}\rangle$ have a similar form.
After that, the whole construction is completely
defined and closed.

Due to the nilpotency of the BRST charge (\ref{Q-final}), solutions to equation \eqref{basic-equation} is
defined up to the gauge transformations
\begin{equation}
\delta|\Phi\rangle=Q|\Lambda\rangle.
\end{equation}
Since $gh(Q)=1$ and $gh(|\Phi\rangle)=0$, the gauge parameter
$|\Lambda\rangle$ has the ghost number $-1$ and therefore its most general
form looks like
\begin{equation}
|\Lambda\rangle=b^+|\lambda^{(s-1)}\rangle,
\end{equation}
where the vector $|\lambda^{(s-1)}\rangle$ has decomposition in creation and annihilation operators similar with
vector $|\phi^{(s)}\rangle$ in (\ref{FockV}).

The equation of motion $Q|\Phi\rangle=0$ in terms of the vectors
$|\phi\rangle,|\phi_1\rangle, |\phi_2\rangle$ can be rewritten as follows\footnote{In what follows, we will often omit superscripts.}
\begin{eqnarray*}
&&p^2|\phi\rangle- l^+|\phi_1\rangle=0\,,
\\
&& p^2|\phi_2\rangle-l|\phi_1\rangle=0\,,
\\
&& l|\phi\rangle-l^+|\phi_2\rangle+(N +\bar{N}+2)|\phi_2\rangle=0\,.
\end{eqnarray*}
In this case, the gauge transformations
$\delta|\Phi\rangle=Q|\Lambda\rangle$ look like
\begin{eqnarray*}
\delta|\phi\rangle=l^+|\lambda\rangle,\quad
\delta|\phi_1\rangle=p^2|\lambda\rangle,\quad
\delta|\phi_2\rangle=l|\lambda\rangle
.
\end{eqnarray*}
The gauge invariant Lagrangian
is constructed as follows
\begin{equation}\label{LagBRST}
{\cal L}=\frac12\int d\theta\langle\Phi|Q|\Phi\rangle.
\end{equation}

It is not difficult to rewrite the Lagrangian (\ref{LagBRST}) in explicit component form.
Putting the expansion (\ref{FockTV}) into the Lagrangian
(\ref{LagBRST}) and integrating over ghost $\theta$ according to the rule
\begin{equation}
\int d\theta\langle 0|\theta|0\rangle=1, \quad \int d\theta\langle
0||0\rangle=0,
\end{equation}
one obtains
\begin{eqnarray}
\label{Lagr}
{\cal L}&=&\frac12\Bigl\{\langle\phi| (p^2|\phi\rangle-
l^+|\phi_1\rangle)\nn\\
&&-\langle\phi_1| (l|\phi\rangle-l^+|\phi_2\rangle+(N
+\bar{N}+2)|\phi_1\rangle)\nn\\
&&-\langle\phi_2|(p^2|\phi_2\rangle-l|\phi_1\rangle)
\Bigr\}.
\end{eqnarray}
Now, using the relation (\ref{FockV}) and the analogous relations for $|\phi_1\rangle$ and
$|\phi_2\rangle$, we define the new Fock space vectors $|H\rangle,\,|C\rangle,\,|D\rangle$ of the form
\begin{eqnarray}
|\phi\rangle&=&|H\rangle=\frac{1}{s!}H_{a(s)}{}^{\da(s)}c{}^{a(s)}c{}_{\da(s)}|0\rangle,
\\
|\phi_1\rangle&=&|C\rangle=\frac{1}{(s-1)!}C_{a(s-1)}{}^{\da(s-1)}c{}^{a(s-1)}c{}_{\da(s-1)}|0\rangle,
\\
|\phi_2\rangle&=&|D\rangle=\frac{1}{(s-2)!}D_{a(s-2)}{}^{\da(s-2)}c{}^{a(s-2)}c{}_{\da(s-2)}|0\rangle.
\end{eqnarray}
Then for given spin $s$, the Lagrangian (\ref{Lagr}) takes the form
\begin{eqnarray}\label{LagCom}
{\cal L}&=&\frac12\Bigl\{
H_{a(s)}{}^{\da(s)}(\partial^2 H^{a(s)}{}_{\da(s)}-{s}\partial^a{}_{\da}C^{a(s-1)}{}_{\da(s-1)})\nn\\
&&-C_{a(s-1)}{}^{\da(s-1)}
(-{s}\partial_b{}^{\db}H^{a(s-1)b}{}_{\da(s-1)b}
-{(s-1)}\partial^a{}_{\da}D^{a(s-2)}{}_{\da(s-2)}+2sC^{a(s-1)}{}_{\da(s-1)})\nn\\
&&-D_{a(s-2)}{}^{\da(s-2)}(\partial^2
D^{a(s-2)}{}_{\da(s-2)}+{(s-1)}\partial_b{}^{\db}C^{a(s-2)b}{}_{\da(s-2)b})
\Bigr\}.
\end{eqnarray}
The corresponding gauge transformations are written as follows
\begin{eqnarray}\label{GTCom}
\delta H_{a(s)}{}^{\da(s)}
&=&
\frac{1}{s}\partial_{a}{}^{\da}\lambda_{a(s-1)}{}^{\da(s-1)}
\nn
\\
\delta C_{a(s-1)}{}^{\da(s-1)}
&=&
\partial^2\lambda_{a(s-1)}{}^{\da(s-1)}
\\
\delta D_{a(s-2)}{}^{\da(s-2)}
&=&
-(s-1)\partial^b{}_{\db}\lambda_{a(s-1)b}{}^{\da(s-1)\db}\nn.
\end{eqnarray}
One can show that Lagrangian \eqref{LagBRST} describes massless spin
$s$ field \cite{Buchbinder:2015kca,Buchbinder:2018yoo}. Indeed,
after removing field $C$ from Lagrangian \eqref{LagCom} with the
help of its equation of motion we are left with two traceless fields
$H$ and $D$. These two traceless fields can be combined into one
double traceless field and Lagrangian for this double traceless
field will coincide with the Fronsdal's Lagrangian
\cite{Fronsdal:1978rb}.

\section{Construction of cubic interaction}

For deriving the cubic interactions we use three copies of
the vectors in extended Fock space $|\Phi_i\rangle, i=1,2,3$ and
three corresponding operators. These operators satisfy the commutation
relations
\begin{equation}
[a_i^a,c_j{}^b]=\delta_{ij}\varepsilon^{ab},\qquad
[a_i^\da,c_j{}^\db]=-\delta_{ij}\varepsilon^{\da\db}
\,,
\end{equation}
\begin{equation}
\{\theta_i,\pi_j\}=\{c_i,b_j^+\}=\{c_i^+,b_j\}=\delta_{ij}
\,.
\end{equation}
The full interacting Lagrangian up to cubic level is defined as follows
\begin{equation}\label{LagBRST_Int}
{\cal L}=\frac{1}{2}\sum_i\int d\theta_i\langle\Phi_i|Q_i|\Phi_i\rangle
+
\frac{1}{2}g\int
d\theta_1d\theta_2d\theta_3\langle\Phi_1|\langle\Phi_2|\langle\Phi_3||V\rangle+h.c.,
\end{equation}
where $|V\rangle$ is some cubic vertex, which should be found, and $g$ is a coupling
constant. It is easy to check that Lagrangian (\ref{LagBRST_Int}) is invariant under
the following gauge transformations up to $g^2$ terms (in what follows
$i\simeq i+3$)
\begin{eqnarray}
\delta|\Phi_i\rangle&=&Q_i|\Lambda_i\rangle-g\int
d\theta_{i+1}d\theta_{i+2}(\langle\Phi_{i+1}|\langle\Lambda_{i+2}|+
\langle\Phi_{i+2}|\langle\Lambda_{i+1}|)|V\rangle,
\end{eqnarray}
if the following condition takes place
\begin{eqnarray}\label{BRST_Cond}
\hat{Q}|V\rangle=0\,,
\qquad
\hat{Q}=\sum_{i=1}^3 Q_i.
\end{eqnarray}
This condition is considered as an equation for $|V\rangle$.
To guarantee zeroth ghost number of Lagrangian (\ref{LagBRST_Int}), the vertex
$|V\rangle$ must have ghost number 3. We will looking for vertex in
the form
\begin{eqnarray}\label{VertexG}
|V\rangle=V|\Omega\rangle,\quad
|\Omega\rangle=\theta_1\theta_2\theta_3|0_1\rangle\otimes|0_2\rangle\otimes|0_3\rangle,
\end{eqnarray}
where the operator $V$ has the ghost number $0$ and depends on
operators $c_i^a,c_i^\da,c_i^+,b_i^+,\pi_i$ as well as on momenta
$p_i^{a\da}$ satisfying the momenta conservation condition
\begin{eqnarray}
\sum_ip^{a\da}_i=0.
\label{p=0}
\end{eqnarray}
However, the equation (\ref{BRST_Cond}) do not determine the vertex
$|V\rangle$ uniquely. Indeed if vertex $|V\rangle$ satisfies the
equation (\ref{BRST_Cond}) then the vertex
\begin{eqnarray}\label{RF}
|V\rangle=|V\rangle+\hat{Q}|W\rangle
\end{eqnarray}
also satisfies this equation. Vertices of the form
$\hat{Q}|W\rangle$ (BRST exact) can be obtained from the free theory
via field redefinitions
\begin{equation}\label{FR}
|\Phi_i\rangle\rightarrow |\tilde\Phi_i\rangle=|\Phi_i\rangle+\int
d\theta_{i+1}d\theta_{i+2}\langle\Phi_{i+1}|\langle\Phi_{i+2}|W\rangle.
\end{equation}
Here $|W\rangle$ is a vector with ghost number 2. Our aim is to find
a operator $V$ in (\ref{VertexG}) which satisfies the BRST
invariance condition (\ref{BRST_Cond}) and determined up to
transformation (\ref{RF})  and can not be removed by the field
redefinition \eqref{FR}. We will call such vertices as BRST-closed.
One can use ambiguity \eqref{RF} in the solution to equation
(\ref{BRST_Cond}) to obtain different explicit forms of the same
physical vertex.

\section{Solutions to the cubic vertices $V$}

First of all, we note that there are six operators $S_i$,
$\bar{S}_i$ \eqref{SbarS} commuting with the BRST operator
\eqref{BRST_Cond}. As a consequence of this we can decompose the
vertex $|V\rangle$ as
\begin{eqnarray}
|V\rangle&=&\sum_{s_i=0}^\infty |V(s_1,s_2,s_3)\rangle
\\
&&
S_i|V(s_1,s_2,s_3)\rangle=\bar{S}_i|V(s_1,s_2,s_3)\rangle=s_i|V(s_1,s_2,s_3)\rangle
\label{S_i_V}
\end{eqnarray}
and solve equation on the vertex \eqref{BRST_Cond} for each values of the spins $s_i$ separately
\begin{equation}\label{BRST_Cond_S}
\hat{Q}|V(s_1,s_2,s_3)\rangle=0.
\end{equation}

Secondly, we note that if we want to construct an interaction for
fields with spins $s_i$ then the operator $V(s_1,s_2,s_3)$
\eqref{VertexG} must obligatory have terms without the ghost fields,
i.e. terms constructed only from the operators $c_i^a$, $c_i^\da$
and the momenta $p_i^{a\da}$. The terms of the operator
$V(s_1,s_2,s_3)$ with the ghost fields are found from equation
\eqref{BRST_Cond_S}.

Thirdly, the terms of the operator $V(s_1,s_2,s_3)$ without the
ghost fields must not depend on the operators $p_i^2$ and $l_i^+$
since these operators are contained in the BRST operators $Q_i$
\eqref{Q-final} and as a consequence such terms can be removed from
the vertex operator with the help of transformation \eqref{RF}.

Let us turn to finding explicit solutions to equation \eqref{BRST_Cond_S}.

Taking into account the remarks made above and condition \eqref{S_i_V} one can show that for three scalar fields we get the vertex operator $V(0,0,0)=const$.

Next let us consider the case $s_1=1$, $s_2=s_3=0$. In this case the
part of the operator $V(1,0,0)$ without the ghost fields must be
linear both in $c_1^a$ and in $c_1^\da$ and due to remarks made
above and due to condition \eqref{p=0} the only possible operator is
$c_1^{a}(p_{2}-p_{3}){}_{a\da}c_1^{\da}$. The rest part of the
operator $V(1,0,0)$ which depends on the ghost fields are found from
equation \eqref{BRST_Cond_S}. The final result for the operator
$V(1,0,0)$ is
\begin{eqnarray}
V(1,0,0)\equiv{}L_1&=&c_1^{a}c_1^{\da}(p_{2}-p_{3}){}_{a\da}
-2c_1^+(\pi_{2}-\pi_{3})
\,.
\end{eqnarray}
Similar expression was found in \cite{Metsaev12}, but in our case
$[\hat{Q}, L_1]\ne0$. Vanishing of  commutator $[\hat{Q}, L_1]$ in
\cite{Metsaev12}  is a consequence that the tracelessness constraint
was not taken into account in the BRST charge in \cite{Metsaev12}.
Nonetheless we try to generalize the result of \cite{Metsaev12} to
our case, namely, we will looking for a solution for the vertex
operator $V(s_1,0,0)$ in a similar form
\begin{eqnarray}
V(s_1,0,0)\equiv{}{\mathbb L}_1^{(s_1)}
&=&
L_1^{s_1}+\text{terms proportional to the ghost fields}
\,.
\end{eqnarray}
Doing so, we find
\begin{eqnarray}
{\mathbb L}_1^{(s_1)}
&=&
L_1^{s_1}
+
s_1(s_1-1)L_1^{s_1-2}c_1^+
\Bigl[l_1^+(2\pi_{2}+2\pi_{3}-\pi_1)-2L_1(\pi_{2}-\pi_{3})\Bigr]
\,.
\end{eqnarray}

Vertex operators $V(0,s_2,0)\equiv{}{\mathbb L}_2^{(s_2)}$ and $V(0,0,s_3)\equiv{}{\mathbb L}_3^{(s_3)}$ have analogous form
\begin{eqnarray}
{\mathbb L}_i^{(s_i)}
&=&
L_i^{s_i}
+
s_i(s_i-1)L_i^{s_i-2}c_i^+
\Bigl[l_i^+(2\pi_{i+1}+2\pi_{i+2}-\pi_i)-2L_i(\pi_{i+1}-\pi_{i+2})\Bigr]
\,,
\\
&&
L_i
=
c_i^{a}c_i^{\da}(p_{i+1}-p_{i+2}){}_{a\da}
-2c_i^+(\pi_{i+1}-\pi_{i+2})
\,.
\end{eqnarray}

Since for $i\ne j$
\begin{eqnarray}
[[\hat{Q},{\mathbb L}_i^{(s_i)}],{\mathbb L}_{j}^{(s_j)}]
=0\,,
\qquad
[\hat{Q},{\mathbb L}_i^{(s_i)}]|\Omega\rangle=0
\end{eqnarray}
then we can construct a vertex for arbitrary values of spins $s_i$ with the number of derivatives $k_{max}=s_1+s_2+s_3$
\begin{eqnarray}
V(s_1,s_2,s_3;k_{max})|\Omega\rangle
=
{\mathbb L}_1^{(s_1)}{\mathbb L}_2^{(s_2)}{\mathbb L}_3^{(s_3)}|\Omega\rangle
\,.
\end{eqnarray}

Thus the vertex operator $V(s_1,s_2,s_3;k_{max})$ for arbitray values of spins $s_i$ with the number of derivatives $k_{max}$ is found.

\section{Example of higher spin interaction}\label{Section_5}

Various problems of interaction of higher spin fields with scalar fields were considered in many papers (see e.g. \cite{Fotopoulos:2007yq,Bekaert:2009ud,Zinoviev:2010cr}).

Let us consider an explicit example of cubic interaction of one real
massless field with arbitrary spin $s$
\begin{equation}
|\Phi_3\rangle=|H\rangle+\theta_3 b_3^+|C\rangle+c_3^+
b_3^+|D\rangle
\end{equation}
where
\begin{eqnarray}
|H\rangle&=&\frac{1}{s!}H_{a(s)}{}^{\da(s)}c{}^{a(s)}c{}_{\da(s)}|0\rangle,
\\
|C\rangle&=&\frac{1}{(s-1)!}C_{a(s-1)}{}^{\da(s-1)}c{}^{a(s-1)}c{}_{\da(s-1)}|0\rangle,
\\
|D\rangle&=&\frac{1}{(s-2)!}D_{a(s-2)}{}^{\da(s-2)}c{}^{a(s-2)}c{}_{\da(s-2)}|0\rangle.
\end{eqnarray}
and two real massless scalar fields
\begin{eqnarray}
|\Phi_1\rangle=\varphi_1|0\rangle,\quad
|\Phi_2\rangle=\varphi_2|0\rangle.
\end{eqnarray}
The total Lagrangian has form
\begin{equation}
{\cal L}={\cal L}_{free}+{\cal L}_{int}.
\end{equation}
Here ${\cal L}_{free}$ is the free Lagrangian for our system of
fields (\ref{LagCom})
\begin{eqnarray*}
{\cal L}_{free}&=&\frac12\Bigl\{
\varphi_1\partial^2 \varphi_1+\varphi_2\partial^2 \varphi_2\\
&&+H_{a(s)}{}^{\da(s)}(\partial^2 H^{a(s)}{}_{\da(s)}-{s}\partial^a{}_{\da}C^{a(s-1)}{}_{\da(s-1)})\\
&&+C_{a(s-1)}{}^{\da(s-1)}
({s}\partial_b{}^{\db}H^{a(s-1)b}{}_{\da(s-1)b}
+{(s-1)}\partial^a{}_{\da}D^{a(s-2)}{}_{\da(s-2)}-2sC^{a(s-1)}{}_{\da(s-1)})\\
&&-D_{a(s-2)}{}^{\da(s-2)}(\partial^2
D^{a(s-2)}{}_{\da(s-2)}+{(s-1)}\partial_b{}^{\db}C^{a(s-2)b}{}_{\da(s-2)b})
\Bigr\}
\end{eqnarray*}
The interacting Lagrangian ${\cal L}_{int}$ corresponds to the
vertex $V(0,0,s;s)={\mathbb L}_3^{(s)}|\Omega\rangle$
(\ref{LagBRST_Int})
\begin{eqnarray}
{\cal L}_{int}&=&\frac12\,g\int
d\theta_1d\theta_2d\theta_3\langle\Phi_1|\langle\Phi_2|\langle\Phi_3|{\mathbb
L}_3^{(s)}|\Omega\rangle+h.c.,
\end{eqnarray}
where
\begin{eqnarray}
{\mathbb
L}^{(s)}_i&=&L^{s}_i+s(s-1)L_i^{s-2}c_i^+[l_i^+(2\pi_{i+1}+2\pi_{i+2}-\pi_i)-2L_i(\pi_{i+1}-\pi_{i+2})],
\\
L_i&=&c_i^{a}c_i^{\da}(p_{i+1}-p_{i+2}){}_{a\da}
-2c_i^+(\pi_{i+1}-\pi_{i+2}).
\end{eqnarray}
After rewriting this Lagrangian component form, one gets
\begin{eqnarray}\label{LICur}
{\cal L}_{int}&=&
(-1)^{s+1}s!\Bigl\{
H^{a(s)}{}_{\da(s)}
j_{a(s)}{}^{\da(s)} -{(s-1)}
\partial{}_{a}{}^{\da}C^{a(s-1)}{}_{\da(s-1)}
j_{a(s-2)}{}^{\da(s-2)}
\Bigr\},
\end{eqnarray}
where $j_{a(s)}{}^{\da(s)}$ are the higher spin currents constructed
from two scalar fields
\begin{eqnarray}
j_{a(s)}{}^{\da(s)}=\varphi_1(\overrightarrow{\partial}_a{}^\da-\overleftarrow{\partial}_a{}^\da)^s\varphi_2=\sum_{k=0}^sC^k_s(-\partial{}_{a}{}^{\da})^{s-k}\varphi_1(\partial{}_{a}{}^{\da})^k\varphi_2,
\quad
C^k_s=\frac{s!}{k!(s-k)!}
\label{current-s00}
\end{eqnarray}
The relevant gauge transformations for higher spin fields remain as
in free theory
\begin{eqnarray}\label{GTCur}
\delta H_{a(s)}{}^{\da(s)}&=&
\frac{1}{s}\partial_{a}{}^{\da}\lambda_{a(s-1)}{}^{\da(s-1)}, \nn\\
\delta C_{a(s-1)}{}^{\da(s-1)}&=&
\partial^2\lambda_{a(s-1)}{}^{\da(s-1)},
\\
\delta D_{a(s-2)}{}^{\da(s-2)}&=&
-(s-1)\partial^b{}_{\db}\lambda_{a(s-1)b}{}^{\da(s-1)\db}.\nn
\end{eqnarray}
However, the general approach leads to gauge transformations for
scalars
\begin{eqnarray}\label{GTCur1}
\delta\varphi_1&=&(-1)^{s}\;2s!\;g
\Bigl[s\sum_{k=0}^{s-1}C^k_{s-1}(\partial_{a}{}^{\da})^{k}\lambda^{a(s-1)}{}_{\da(s-1)}
(2\partial{}_{a}{}^{\da})^{s-k-1}\varphi_2\nn\\
&&-(s-1)\sum_{k=0}^{s-2}C^k_{s-2}(\partial_{a}{}^{\da})^{k+1}\lambda^{a(s-1)}{}_{\da(s-1)}
(2\partial{}_{a}{}^{\da})^{s-k-2}\varphi_2\Bigr],
\end{eqnarray}
\begin{eqnarray}\label{GTCur2}
\delta\varphi_2&=&2s!\;g
\Bigl[s\sum_{k=0}^{s-1}C^k_{s-1}(\partial{}_{a}{}^{\da})^{k}\lambda^{a(s-1)}{}_{\da(s-1)}
(2\partial_{a}{}^{\da})^{s-k-1}\varphi_1\nn\\
&&-(s-1)\sum_{k=0}^{s-2}C^k_{s-2}(\partial{}_{a}{}^{\da})^{k+1}\lambda^{a(s-1)}{}_{\da(s-1)}
(2\partial_{a}{}^{\da})^{s-k-2}\varphi_1
\Bigr].
\end{eqnarray}

Thus we have explicitly constructed a vertex corresponding to the interaction of a field spin $s$ with two real scalars and deformation of the gauge transformation.

Let us remind once again that there is arbitrariness \eqref{RF} in the explicit form of the interaction vertex and we can use it to get more convenient
 expressions for interaction \eqref{LICur} and/or gauge transformations \eqref{GTCur1} and \eqref{GTCur2}.

\section{Summary}

In this paper we have analyzed and constructed the Lorentz covariant
cubic interactions for completely unconstrained massless higher spin
fields in $d=4$ Minkowski space with the maximum number of derivatives. The construction is given in the
framework of the BRST approach to higher spin fields adopted to
multispinor formalism. Unlike the previous work \cite{BR21}, in the
present formulation there is no need to use the tracelessness constraint for
irreducible massless higher spin fields since we use spin-tensors with dotted and
undotted indices and this constraint is
fulfilled identically what in some sense simplify an analysis.
However, the corresponding BRST operator has a different structure
than that in \cite{BR21}, and the derivation of a cubic vertex now
requires a separate analysis. Such an analysis was given in this
paper.

Within of the BRST approach, the problem of constructing cubic
interaction vertices is reduced to finding a vector $|V\rangle$
(\ref{VertexG}) which should be BRST-closed $\hat{Q}|V\rangle=0$
\eqref{BRST_Cond}. We have carried out a general analysis of the
equation for the cubic vertex and described a procedure of its
finding. For three given massless fields with spins $s_1,s_2,s_3$ we
have constructed a cubic vertex with $k_{max}=s_1+s_2+s_3$ numbers
of derivatives. The case of constructing a vertex with $k_{min}$
number of derivatives will be considered in a future paper.

An explicit example of cubic interaction of a field with spin $s$ with two
real massless scalar fields was constructed in details. The interacting Lagrangian and
the gauge transformations in explicit component form are given by
(\ref{LICur}) and (\ref{GTCur}), (\ref{GTCur1}), (\ref{GTCur2}).

It is evident that the BRST approach to constructing the interacting
vertices for $4d$ completely irreducible higher spin fields in terms
of spin-tensor fields with dotted and undotted indices can be used
for finding the manifestly Lorentz-covariant cubic and higher
vertices for various bosonic and fermionic, massive and massless
higher spin fields. We hope to study all these issues in the
forthcoming papers.

\section{Acknowledgements}
The authors are grateful to R.R. Metsaev, A.A. Reshetnyak and Yu.M.
Zinoviev for useful discussions and comments. The work is supported
by the Ministry of Education of the Russian Federation, project
FEWF-2020-0003.

\end{document}